\input harvmac
\input epsf

\def\l{\lambda}
\def\a{\alpha^{\prime}}
\def\t2{\tau_2}
\def\AdSS5{$AdS_5$}
\def\AdS5s5{$AdS_5 \times S^5$}
\def\gy{g_{{_YM}} }

\def\Hy{\calH^{^{YM}}}

\def\RR{{$R\otimes R$}}

\def\calR{{\cal R}}

\def\calN{{\cal N}}

\def\calH{{\cal H}}

\def\gy{g^2_{_{YM}}}
\def\lr { \lref}
\def\np#1#2#3{Nucl. Phys. {\bf B#1} (#2) #3}
\def\pl#1#2#3{Phys. Lett. {\bf B#1} (#2) #3}

\def\prd#1#2#3{Phys. Rev. {\bf D#1} (#2) #3}

\def\IZ{\relax\ifmmode\mathchoice
{\hbox{\cmss Z\kern-.4em Z}}{\hbox{\cmss Z\kern-.4em Z}}
{\lower.9pt\hbox{\cmsss Z\kern-.4em Z}} {\lower1.2pt\hbox{\cmsss
Z\kern-.4em Z}}\else{\cmss Z\kern-.4em Z}\fi}

\lr\brezin{ E. Brezin and Zinn-Justin,
{\it Renormalization group approach to matrix models},  hep-th/9206035;
Phys. Lett. {\bf B288} (1992) 54.}
\lr\hardy{G.  Hardy, {\it Divergent series}, OUP, 1949.}
\lr\bender{C.M.  Bender and S.A.  Orszag, {\it Advanced mathematical
methods for scientists and engineers}, McGraw--Hill, New York, 1978.}
\lr\zinnjustin{J.C.  Le Guillou and J.  Zinn--Justin (Eds.), {\it
Large-order behaviour of perturbation theory}, North Holland,
Amsterdam, 1990.}
\lr\doreythree{N. Dorey,
T. J. Hollowood,  V. V. Khoze,  M. P. Mattis,  S. Vandoren,
 {\it Multi-Instanton Calculus and the AdS/CFT Correspondence in
  N=4 Superconformal Field Theory}, hep-th/9901128.}
\lr\grossperiwal{D.J.  Gross and V.  Periwal, {\it String perturbation theory diverges},
 Phys. Rev. Lett. 61 (1988)1517.}
\lr\greencorfu{M.B.  Green, {\it Interconnections between string
theory, M theory and $\calN =4$ Yang--Mills}, Proceedings of TMR
meeting, Corfu (September 1998), hep-th/9903124.}
\lr\shenker{S.  Shenker, {\it The strength  of nonperturbative effects in string theory},  Cargese Workshop on Random Surfaces,
 Quantum Gravity and Strings, Cargese, France, May 28 - June 1, 1990.}
\lr\thooft{G. 't Hooft, {\it A planar diagram theory for strong
interactions}, Nucl. Phys. {\bf B72} (1974) 461.}
\lr\berkovaf{N. Berkovits and C. Vafa, {\it Type IIB $R^4 H^{4g-4}$
Conjectures}, hep-th/9803145; Nucl. Phys. {\bf B533}
(1998).}
\lr\berkostrings{N.  Berkovits, {\it Generalization of the $R^4$ conjecture}, Proceedings of Strings'98,
\hfill\break\noindent
 http://www.itp.ucsb.edu/online/strings98/berkovits.}
\lr\tseytruss{J. Russo and  A.A. Tseytlin, {\it One-loop
four-graviton amplitude in eleven-dimensional supergravity}, hep-th/970713;
 Nucl. Phys. {\bf  B508}(1997)245.}
\lr\russo{J.G. Russo, {\it An ansatz for a non-perturbative
four-graviton amplitude in type IIB superstring
     theory}, hep-th/9707241; Phys. Lett. {\bf B417} (1998) 253.}
\lr\bgkr{M.  Bianchi, M.B.  Green, S.  Kovacs and G.C.  Rossi, {\it
Instantons in supersymmetric Yang--Mills and D-instantons in IIB
superstring theory}, hep-th/9807033;  JHEP {\bf 9808} (1998) 13.}
\lr\bg{T.  Banks and M.B.  Green, {\it Nonperturbative effects in
 $AdS_5\times  S^5$ string theory and d = 4 SUSY
Yang-Mills}, hep-th/9804170;  JHEP {\bf 9805} (1998) 2.}
\lr\greensethi{M.B.  Green and S. Sethi,
{\it Supersymmetry constraints on type IIB supergravity}, hep-th/9808061.}
\lr\polchinski{J.  Polchinski, {\it Tasi lectures on D-branes},
 hep-th/9611050.}
\lr\doreyone{N.  Dorey, T.J.  Hollowood, V.V.  Khoze and M.P.  Mattis,
 {\it Multiinstantons and Maldacena's
conjecture}, hep-th/9810243.}
\lr\doreytwo{N.  Dorey, T.J.  Hollowood, V.V.  Khoze,
 M.P.  Mattis and S. Vandoren, {\it Yang-Mills instantons in the large
 N limit and the AdS / CFT correspondence},
 hep-th/9808157.}
\lr\rvb{N. Berkovits and C. Vafa, {\it N=4 Topological Strings},
hep-th/9407190, \np{433}{1995}{123}.}
 \lr\rvo{H. Ooguri and C.
Vafa, {\it All Loop N=2 String Amplitudes}, hep-th/9505183,
\np{451}{1995}{121}.}
\lr\rstrings{N. Berkovits, {\it
Generalization of the $R^4$ conjecture},
 proceedings of  Strings '98\semi
http://www.itp.ucsb.edu/online/strings98/berkovits/.}
 \lr\rWSD{E.
Witten, {\it String Theory Dynamics in Various Dimensions},
hep-th/9503124, \np{443}{1995}{85}.}
 \lr\rPKT{P. K. Townsend, {\it
The Eleven-dimensional Supermembrane Revisited}, hep-th/9501068,
\pl{350}{1995}{184}.}
\lr\rkirit{ B. Pioline and E. Kiritsis,
{\it On $R^4$ Threshold Corrections in IIB String Theory and $(P,
Q)$ String Instantons}, hep-th/9707018, \np{508}{1997}{509}.}
\lr\rpioline{ B. Pioline, {\it A Note On Nonperturbative $R^4$
Couplings}, hep-th/9804023.}
 \lr\rberk{N. Berkovits, {\it
Construction of $R^4$ Terms in N=2 D=8 Superspace},
hep-th/9709116, \np{514}{1998}{191}. }
\lr\russo{J. G. Russo,
{\it Construction of SL(2,Z) Invariant Amplitudes in Type IIB
Superstring Theory}, hep-th/9802090; {\it An Ansatz For A
Nonperturbative Four Graviton Amplitude In Type IIB Superstring
Theory}, hep-th/9707241, \pl{417}{1998}{253}.}
 \lr\kehagias{A.
Kehagias and H. Partouche, {\it The Exact Quartic Effective
Action
 For The Type IIB Superstring}, hep-th/9710023, \pl{422}{1998}{109}.  }
\lr\terras{ A. Terras, {\it Harmonic Analysis on Symmetric Spaces
and Applications I}, Springer-Verlag (New York) 1985. }
\lr\berkovafa{N. Berkovits and C. Vafa, {\it Type IIB $R^4
H^{4g-4}$ Conjectures}, hep-th/9803145. }
\lr\rmIIB{P. Aspinwall,
{\it Some Relationships Between Dualities in String Theory},
hep-th/9508154, Nucl. Phys. Proc. Suppl. {\bf 46} (1996) 30\semi
J. Schwarz, {\it The Power of M Theory}, hep-th/9510086,
\pl{367}{1996}{97}. }
\lr\cjs {E. Cremmer, B. Julia and J.
Scherk, {\it Supergravity Theory in Eleven Dimensions}, Phys.
Lett. {\bf 76B} (1978) 409.}
 \lr\minasduff{M.J.~Duff, J.T.~Liu and
R.~Minasian, {\it Eleven-dimensional Origin of String-String
Duality: A One Loop Test}, hep-th/9506126, Nucl.  Phys. {\bf
452B} (1995) 261.}
\lr\wittbaryon{E.  Witten, {\it Baryons and branes in anti-de Sitter space},
hep-th/9805112;  JHEP 9807:006,1998.}
\lr\vafawitt{C.~Vafa and E.~Witten, {\it A One
Loop Test of String Duality},
  hep-th/9505053, Nucl.  Phys. {\bf B447} (1995) 261.}
\lr\sethia{S.  Sethi,  S.  Paban and M.  Stern, {\it Constraints
{}From Extended Supersymmetry in Quantum Mechanics},
hep-th/9805018; {\it Supersymmetry and Higher Derivative Terms in
the Effective Action of Yang--Mills Theories},  hep-th/9806028.}
\lr\stromingerb{A. Strominger, {\it Loop Corrections to the
Universal Hypermultiplet}, hep-th/9706195.}
\lr\ferrarab{I.
Antoniadis, S. Ferrara, R. Minasian and K.S. Narain, {\it $R^4$
couplings in M and type II theories}, hep-th/9707013.}
\lr\knn{J.Koplik, A.Neveu, S.Nussinov,  {\it Some aspects of the
planar perturbation series}, Nucl. Phys.  {\bf B123} (1977) 109.}
\lr\ew{E.Witten,  {\it Current algebra theorems for the U(1) \lq
Goldstone boson'}, Nucl. Phys.  {\bf B156}  (1979)  269; {\it
Instantons, the quark model and the $1/N$ expansion}, Nucl. Phys.
{\bf B149} (1979) 285.}
\lr\jr{R.Jackiw, C.Rebbi, Phys. Rev. {\bf
D14}, (1976), 517.}
 \lr\hh{G.Horowitz, H.Ooguri, hep-th/9802116.}
\lr\kslvz{S.Kachru, E.Silverstein, {\it 4-D conformal Theories
and Strings on Orbifolds}, hep-th/9802183; A.Lawrence,
N.Nekrasov, C.Vafa, {\it On Conformal Field Theories in Four
Dimensions}, hep-th/9803015. }
 \lr\thorn{C.B.  Thorn, in Sakharov Conference on
Physics, Moscow, (91),447.}
\lr\dpsgketc{S.Gubser, I.Klebanov,
A.Peet, Phys. Rev. {\bf D54}, (1996), 3915, hep-th/9602135 ;
 A.A.Tseytlin, I.Klebanov, Nucl. Phys. {\bf B475}, (1996), 179, hep-th/9604166
;
 S.Gubser, I.Klebanov, Phys. Lett. {\bf B413}, (1997), 41, hep-th/9708005; and
references cited in \juan.}
\lr\maldacena{J.  Maldacena, {\it   The
large $N$ limit of superconformal field theories and
supergravity}, hep-th/971120.}
\lr\bvz{M.Bershadsky,
Z.Kakushadze, C.Vafa, {\it String Expansion as Large N Expansion
of Gauge Theories}, hep-th/9803076; M.Bershadsky, A.Johansen,
{\it Large N Limit of Orbifold Field Theories}, hep-th/9803249.}
\lr\gk{S.S. Gubser and I.R.  Klebanov, {\it Absorption by branes
and Schwinger terms in the world volume theory}, hep-th/9708005.}
\lr\gkp{S.S. Gubser,  I.R.  Klebanov and A.M.  Polyakov, {\it
Gauge theory correlators from non-critical string theory},
hep-th/9802109.}
\lr\wittone{E.  Witten, {\it Anti de Sitter
Space and Holography}, hep-th/9802150.}
\lr\tHS{G. 't Hooft, {\it
Dimensional Reduction in Quantum Gravity}, gr-qc/9310006;
L.Susskind, {\it The World as a Hologram}, J. Math. Phys. 36,
6377 (1995), hep-th/9409089.}
\lr\gv{M.B.  Green and P. Vanhove,
{\it D-instantons, Strings and M-theory}, Phys. Lett. {\bf B408}
(1997) 122, hep-th/9704145.}
\lr\ggv{M.B.~Green,  M.~Gutperle and
P.~Vanhove, {\it One Loop in Eleven-Dimensions}, hep-th/9706175,
\pl{409}{1997}{177}.}
\lr\ggk{M.B.~Green,  M.~Gutperle and
H.~Kwon, {\it Sixteen Fermion
 and Related
Terms in M theory on $T^2$}, hep-th/9710151.}
\lr\wittfive{E.
Witten, {\it Five-brane Effective Action in M Theory},
 hep-th/9610234,
 J. Geom. Phys. {\bf 22} (1997) 103.}
\lr\nilsson{B.E.W. Nilsson and A.  Tollsten, {Supersymmetrization
of $\zeta(3) R_{\mu\nu\tau\sigma}^4$ in superstring theories},
Phys. Lett {\bf 181B} (1986) 63.}
\lr\greengut{M.B.~Green and
M.~Gutperle, {\it Effects of D-instantons}, hep-th/9701093,
\np{498}{1997}{195}.}
\lr\greenschwarz{M.B.  Green and J.H.
Schwarz, {\it Supersymmetric Dual String Theory (II).  Loops and
Renormalization}, Nucl.  Phys. {\bf B198} (1982) 441.}
\lr\osborn{ J.  Erdmenger and H.  Osborn, {\it Conformally
covariant differential operators: Symmetric tensor fields},
gr-qc/9708040; Class. Quantum Grav. {\bf 15} (1998) 273.}
\lr\eguchi{T.  Eguchi, {\it S Duality and Strong Coupling
Behavior of Large N Gauge Theories with N=4 Supersymmetry},
hep-th/9804037.}
\lr\schwarzwest{J.H.  Schwarz and P.C.  West,
{\it  Symmetries and Transformations of Chiral $N=2$, $D=10$
Supergravity}, Phys. Lett.
 {\bf 126B} (1983) 301.}
\lr\greenschwarza{M.B.  Green and J.H.  Schwarz, {\it Extended
 Supergravity in Ten Dimensions}, Phys.  Lett.  {\bf 122B} (1983) 143.}
\lr\schwarza{J.H.  Schwarz, {\it Covariant Field Equations of
Chiral $N=2$, $D=10$ Supergravity},  Nucl. Phys.  {\bf B226}
(1993) 269.}
\lr\howest{P.S.  Howe and P.C. West, {\it The
Complete $N=2$ $D=10$ Supergravity}, Nucl.  Phys.  {\bf B238}
(1984) 181.}
\lr\matrixth{T.~Banks, W.~Fischler, S.H.~Shenker and
L.~Susskind, {\it M Theory As A Matrix Model: A Conjecture},
hep-th/9610043, \prd{55}{1997}{5112}.}
\lr\gw{D.J.~Gross and
E.~Witten, {\it Superstring modifications of Einstein's
    equations}, \npb277(1986)1.}
\lr\gris{M.T.~Grisaru , A.E.M~Van de Ven and D.~Zanon, {\it
Two-dimensional supersymmetric sigma models on Ricci flat
K\"ahler manifolds are not finite}, \npb277(1986)388 ; {\it Four
loop divergences for the N=1 supersymmetric nonlinear sigma model
in two-dimensions}, \npb277(1986)409.}
\lr\jackreb{R.  Jackiw and
C. Rebbi, {\it Spinor analysis of Yang--Mills theory}, Phys. Rev.
{\bf D16} (1977) 1052.}
\lr\greengutc{M.B.~Green and M.~Gutperle,
{\it D-particle Bound States and the D-instanton Measure},
hep-th/9711107, {\bf JHEP01} (1998) 005.}
\lr\f{D.Z.Freedman,
S.D.Mathur, A.Matusis, L.Rastelli, {\it Correlation Functions in
the CFT(D)/ADS(D+1) Correspondence}, hep-th/9804058.}
\lr\greencargese{M.B.  Green, {\it Connections between M theory
and superstrings}, Proceedings of the 1997 Advanced Study
Institute on Strings, Branes and Dualities, Cargese,
\hep-th/9712195; Nucl. Phys. Proc. Suppl. {\bf 68} (1998) 242.}
\lr\moorenek{G.  Moore, N.  Nekrasov, S. Shatashvili, {\it D particle
bound states and generalized instantons},  hep-th/9803265.}
\lr\nicolaia{W.  Krauth, H.  Nicolai, Staudacher, {\it Monte Carlo
approach to M theory}, hep-th/9803117;  Phys. Lett. {\bf B431}
(1998) 31.}
\lr\kosvan{I.K.  Kostov and P.  Vanhove, {\it Matrix string partition
functions}, hep-th/9809130; Phys. Lett. {\bf B444} (1998) 196.}
\lr\gopa{R. Gopakumar and C.  Vafa, {\it On the Gauge Theory/Geometry
Correspondence},  hep-th/9811131.}
\lr\johnson{C.V.  Johnson, {\it Etudes on D-branes}, hep-th/9812196.}
\lr\thf{G. 't Hooft, ``On the Convergence of Planar Diagram Expansions'',
 Commun.Math.Phys. {\bf 86}, (1982), 449.}
\lr\nonren{H.  Osborn (private communication);
S.~Gubser, I.~Klebanov, {\it Absorption by Branes and Schwinger
Terms in the World Volume Theory}, hep-th/9708042.
Phys. Lett. {\bf 413B}, 41 (1997);
D.~Anselmi, D.Z.~Freedman, M.T.~Grisaru,
A.A.~Johansen, {\it Nonperturbative Formulas for Central Functions of
Supersymmetric Gauge Theories}, hep-th/970804;
S.~Lee,~S.~Minwalla,~M.~Rangamani,~and~N.~Seiberg,
{\it Three-Point Functions of Chiral Operators in $D=4$~,~${\cal N}=4$ SYM at
Large $N$},  hep-th/9806074;
 E.~D'Hoker,~D.Z.~Freedman,~and~W.~Skiba,
{\it Field Theory Test for Correlators in the AdS/CFT
Correspondence}, hep-th/9807098;
B.~Eden, P.S.~Howe, C.~Schubert, E.~Sckatchev, and P.C.~West,
{\it Four-point Functions in ${\cal N}=4$ Supersymmetric Yang-Mills Theory at
Two Loops}, hep-th/9811172.}
\lr\brogut{ J. Brodie, M. Gutperle, {\it
String corrections to four point functions in the AdS/CFT correspondence},
hep-th/9809067, Phys. Lett. {\bf 445B},296 (1999).}
\lr\evashamit{S. Kachru, E. Silverstein, {\it 4d Conformal Field
Theories and Strings on Orbifolds}, hep-th/9802183,
Phys. Rev. Lett. {\bf 80}, 4855, (1998).}
\lr\lnv{A. Lawrence, N. Nekrasov, C. Vafa, {\it On Conformal Theories
in Four Dimensions}, hep-th/9803015,
Nucl.Phys. {\bf B533}, 199, (1998).}

\noblackbox
\Title{\vbox{
\hbox{HUTP-99/A037, DAMTP-1999-98}
\hbox{\tt hep-th/9908020}
}}{Instantons and Non-renormalisation
in AdS/CFT}
\centerline{Rajesh Gopakumar\foot{gopakumr@tomonaga.harvard.edu}}
\smallskip
\centerline{\it Lyman Laboratory of Physics, Harvard University}
\centerline{\it Cambridge, MA 02138, USA}
\vskip 0.05in
\centerline{Michael B. Green\foot{ M.B.Green@damtp.cam.ac.uk}}
\smallskip
\centerline{\it DAMTP, Silver Street, Cambridge CB3
9EW, UK}

\vskip .15in

The series of perturbative fluctuations around a multi-instanton
contribution to  a specific class of correlation functions of supercurrents
in $\calN=4$ supersymmetric $SU(N)$ Yang--Mills theory is examined in the
light of the AdS/CFT correspondence.
Subject to certain  plausible assumptions,  we argue that
a given term in the $1/N$ expansion in
such a background receives only  a finite number of perturbative
corrections in the 't Hooft limit.
Such instanton non-renormalisation theorems would explain,  for example,
the exact agreement of certain weak coupling Yang--Mills
instanton calculations with the strong coupling predictions
arising from D-instanton effects in
string theory amplitudes. These non-renormalisation theorems
essentially follow from the assumption of a well defined derivative $(\a)$
expansion in the string theory dual of the Yang--Mills theory.

\Date{August 1999}

\newsec{Introduction}

The conjectured equivalence of type IIB superstring theory on
$AdS_5\times S^5$ to the
boundary $\calN=4$ supersymmetric $SU(N)$ Yang--Mills conformal
field theory  \refs{\maldacena,\gkp,\wittone}
has been tested by a variety of calculations at
leading order in the large-$N$ limit and at large values of the
't Hooft coupling, $\lambda= \gy N/4\pi$ ($\gy$ is the
Yang--Mills coupling constant). Many of these tests, such as those
of certain two and three point correlation functions, have relied
on non-renormalisation theorems/conjectures \refs{\nonren} and therefore
allow the meaningful comparison of  the regimes of strong and weak 't Hooft
coupling.

Clearly, the ideal way of developing the AdS/CFT correspondence beyond
 the limited large $\l$ region in which it has so far been studied
 would be to explicitly quantize IIB superstring theory in an \AdS5s5\
 background. Unfortunately, this
is a daunting problem, even at tree level -- in part because of
the presence of a nonzero condensate of \RR\ background fields
associated with the nonzero $F_5$ flux.   In the absence of an
explicit construction of string amplitudes most concrete
calculations have made use of known low order terms in the expansion of
the effective supergravity action in powers of the dimensionless
parameter $\alpha'/L^2$
($L$ is the size of the $AdS_5$ and $S^5$ background and
${\alpha'}^{1/2}$ is the string distance scale).
In fact, knowing the complete effective action for the massless fields of string
theory would  be sufficient to  compute the Yang-Mills
correlation functions of the relevant dual operators, but we are far from
 achieving  this.

Nevertheless,
in the following we will show how some reasonable assumptions concerning the
structure
of the low energy expansion
of type IIB string theory lead to a number of non-renormalisation theorems
in the instanton sector of certain Yang-Mills correlation functions.
To be concrete, we will consider the D-instanton
contributions to the four graviton scattering amplitude and arrive at
statements regarding
the corresponding Yang-Mills instanton
terms in the AdS/CFT dual   correlation
functions of four energy-momentum tensors. Similar statements also apply
to any of the Yang-Mills correlations functions that are related by
supersymmetry.
Specifically, what we will see is that, for certain `protected' parts of the
correlation functions, the 't Hooft expansion around an instanton
background has only a finite number of perturbative
terms in $\l$ at each order in $1/N$.
In particular, we will see that at leading order
in $N$  only the $\l$-independent
semi-classical term arises.  This would account for the precise
agreement  (for any instanton number $K$, at leading order
in $N$) between the D-instanton contributions,
at leading order in the  $\alpha'$ (or $1/\l$) expansion
\refs{\bg},  with the semi-classical (small-$\l$)
contributions of Yang--Mills instantons
 \refs{\bgkr,\doreythree}. In the case of
two and three point functions the space-time dependence is completely
determined by (super)
conformal invariance.  However, the matching of the string theory
D-instanton and Yang--Mills
instanton contributions to the protected correlation functions involves
matching non-trivial functions of the space-time positions (functions of
two independent cross ratios in the case of the correlator of four stress
tensors), together with
specific dependence on $N$, $\lambda$ and the instanton number.

\subsec{Overview of the Correspondence}
The AdS/CFT conjecture \refs{\maldacena} gives a relation between
the parameters of the string theory -- the dimensionless \AdS5s5\ scale
 $L^2/\alpha'$,   the \RR\ scalar field,
 $C^{(0)}$, and the coupling constant $g =
e^\phi = \tau_2^{-1}$ -- and those of the Yang--Mills
 theory with gauge group $SU(N)$.
\eqn\mapss{g= {g^2_{_{YM}}\over 4\pi},\qquad  2\pi
C^{(0)}=  \theta , \qquad {L^4\over {\alpha'}^2} =
 g_{_{YM}}^2 N \equiv 4\pi \lambda,}
where $\theta$ is the
constant axionic angle.  This means that the constant value of
 the complex coupling constant, $\tau\equiv \tau_1 + i \tau_2
= C^{(0)} + i e^{-\phi}$,  in the \AdS5s5\ background is
identified with the complex Yang--Mills coupling,
\eqn\comcoup{\tau  = {\theta \over 2\pi} + i
{4\pi \over \gy}.}
 In the following, $\tau$ will always be assumed to be equal to
 this constant value (mostly with $\tau_2^{-1} <<1$).

According to the prescription of \refs{\gkp, \wittone}
the amplitudes of the bulk
 superstring theory in the \AdS5s5\ background with
 fields propagating to specified values at
 points on the boundary are equivalent to
correlation functions of composite operators in the boundary
Yang--Mills theory.  The boundary values of the bulk fields are
interpreted as sources coupling to the operators in the
Yang--Mills theory.
Collectively denoting the
independent cross ratios of the positions of the boundary fields
by $\eta$, the resulting amplitude for an $n$-point function
in the gauge theory can be written as a finite sum over contributions of
the form
\eqn\fouramp{{\cal H}_n^s\left({\a\over L^2}, \tau, {\bar \tau}, \eta
\right)A_n =\calH_n^{YM}(\lambda,N,\theta,\eta)A_n .}
The right-hand  side is just a rewriting in terms of the
variables $N$, $\lambda$ and $\theta$
in which it is natural
to express the correlation functions of the
Yang--Mills theory. The finite sum involves
factors $A_n$ which span an independent set of tensor
structures consistent with the space-time quantum numbers and symmetries
of the $n$-point functions. Moreover, the $A_n$'s all have a common factor,
also dictated by symmetry (a function of the space-time separations
$|x_i-x_j|$ ), which carries the dimension of the correlation
function. Therefore, almost all the non-trivial information about the correlation
functions really lie in the functions ${\cal H}_n$.

In the following, we will be exclusively concerned with the large $N$ limit
of  \fouramp. The expansion of gauge theory amplitudes in $1/N$
translates into a small $g$ expansion in the left-hand side of
\fouramp.
The 't Hooft expansion of Yang--Mills amplitudes takes the familiar form
(for convenience, we will drop the subscript $n$ in much that follows),
\eqn\genn{\Hy(\lambda,N, \theta,\eta) =N^2
\left[ \Hy_0(\lambda,\eta) + {1\over N^2}\Hy_1(\lambda,\eta) + \dots +
{1\over N^{2k}} \Hy_k (\lambda,\eta) +\dots \right  ] +\dots .}
Here the second ellipsis
includes Yang--Mills
instanton terms of the form
$e^{-2\pi |K| {N\over \lambda} + iK\theta}$, each coming
with its series of fluctuations. Although these instanton terms are exponentially
 suppressed they are uniquely specified by their phase. It is the structure
of the 't Hooft expansion of fluctuations
around a particular instanton background that we will focus on in Sec. 2.

In the correspondence with the dual string theory, the terms in \genn\
which are
powers of $1/N$ arise from perturbative string contributions with $k$ being
the world sheet genus. The instanton terms which are suppressed by
powers of $e^{-N}$
are  non-perturbative and can be identified with
D-instanton contributions \refs{\greengut}.
Though the form of the $1/N$ expansion in \genn\ was originally
motivated by weakly coupled
perturbation theory, the existence of the dual string theory with the
identifications \mapss\ implies such a form should apply for all $\l$.
In particular, we will exploit the existence of a well defined expansion
for large $\l$ that is defined by the $\a$ expansion of the string theory.

\newsec{Instanton Non-renormalisation Theorems}

 The instanton calculations in
\refs{\bgkr, \doreythree} involved correlation
functions of various combinations of the superconformal currents that
make up a short (256 component) $\calN=4$ supermultiplet.  However, the
general structure of interest to us does
 not depend on which of these correlation functions is considered so
 we will  focus  on a specific tensor
structure in the
correlation function of four stress
tensors.

This particular tensor structure
can be defined
by its relation, via the AdS/CFT correspondence,
to the ${\cal R}^4$ term in the type
IIB effective action (where $\calR$ denotes the Weyl curvature).
 This  ten-dimensional term has the tensor structure
\eqn\rfour{{\cal R}^4 \equiv t^{M_1\cdots M_8}t_{N_1\cdots N_8}
R^{N_1N_2}_{M_1M_2}\cdots R^{N_7N_8}_{M_7M_8}}
with $t$ being a standard eighth-rank tensor.  We wish to consider the
linearization of the four curvatures around the \AdS5s5\ background,
keeping only the
polarisations in the $AdS_5$ directions.
The four-graviton scattering  amplitude is expressed as a
functional of the boundary ($S^4$) values of the graviton
by attaching a spin-two
bulk-to-boundary propagator to each graviton leg in the linearized vertex.
Since the boundary graviton field is
interpreted  in the  Yang--Mills theory
as the source  for the
stress tensor, this procedure defines a particular tensor
contribution to the correlation  function of
 four stress tensors \refs{\bg} which can be expressed as
\eqn\leadcont{{L^2\over \a}\t2^{1/2} f_1^{(0,0)}(\tau,\bar \tau)
g_1(\eta)A_4.}
The ${\a}^{-1}$ dependence reflects the fact that
the ${\cal R}^4$ term is of order ${\a}^3$ relative to the
Einstein--Hilbert term.   The function
$A_4$ has, in addition to  the particular tensor structure
determined by the bulk-boundary correspondence described above,
a factor of $\prod_{i<j} |x_i-x_j|^{-{8\over 3}}$, which is fixed by
conformal invariance.
The residual dependence on the positions of the
boundary operators is
contained in the function of the two independent cross-ratios which has
been denoted as $g_1(\eta)$. Explicit expressions for
$g_1(\eta)$ in the case of
 closely related four-point functions were obtained
in \refs{\bgkr, \brogut}.
The  modular invariant function of the
(complex)  string coupling,
$f_1^{(0,0)}(\tau,\bar \tau)$, is a nonholomorphic Eisenstein series
that has the Fourier expansion in
powers of $e^{2\pi i \tau_1}$ \refs{\greengut},
\eqn\fone{ \eqalign{   f_1^{(0,0)} (\tau, \bar \tau) & \equiv
 {\sum_{(m,n)\ne(0,0)}}
{\tau_2^{3/2}\over |m+n \tau|^3} =
\sum_{K=-\infty }^\infty {\cal F}^1_K(\tau_2) e^{2\pi i K \tau_1} \cr
&= 2\zeta(3)\tau_2^{{3\over 2}} + {2\pi^2\over 3}\tau_2^{-{1\over
2}} + 4 \pi \sum_{K=1}^\infty |K|^{1/2} \mu(K,1) \cr & \times
\left(e^{2\pi i K \tau} + e^{-2\pi i K \bar \tau} \right) \left(1
+ \sum_{k=1}^\infty (4\pi K \tau_2)^{-k} {\Gamma( k -1/2)\over
\Gamma(- k -1/2) k!} \right) .\cr}}
Here $\mu(K,1)=\sum_{d|K}  d^{-2}$.
The $K=0$ term contains the perturbative tree-level and one-loop
contributions while the $K\ne 0$  terms are D-instanton contributions.
The leading $\tau_2$ independent
term in the charge-$K$ D-instanton sector
was found to agree with a weak coupling Yang-Mills calculation
in \refs{\doreythree} (at least for the related sixteen-dilatino
correlation function).

More generally, in the \AdS5s5\  background there will be contributions
proportional to $A_4$ which are of higher order in $\a$. These come
from higher derivative terms in the effective action and can be studied
in a Taylor expansion for  small $\a/L^2$,
\eqn\taylexp{{\cal H}^s\left({\a\over L^2}, \tau, {\bar \tau}, \eta\right)A_4
= \sum_{l=1}\left({\a\over L^2}\right)^{l-2}F_l(\eta, \tau, {\bar \tau})
\, A_4,}
where the
${\cal R}^4$ contribution is the first ($l=1$) term in the series
(so that $F_1(\eta, \tau, {\bar \tau})=g_1(\eta)\t2^{1/2}
f_1^{(0,0)}(\tau,\bar \tau)$).   Examples of
higher derivative terms in the ten dimensional effective action
that would contribute to \taylexp\ include terms of the general
form (in string frame)
\eqn\stringfull{(\alpha')^{2k-3}\int d^{10}x \,\sqrt{G^{(10)}}
\,e^{(5k-11/2)\phi}\, F_5^{4k-4}\, \calR^4\, f^{(0,0)}_k(\tau,\bar\tau).}
The modular functions $f_k^{(0,0)}$ that appear here, have been conjectured to be
generalised Eisenstein functions \refs{\berkovaf,\berkostrings}. In
the \AdS5s5\ background with its constant five-form field strength $F_5$,
these terms can give contributions proportional to $A_4$.
Another class of terms suggested in \refs{\russo,\tseytruss}
 involve derivatives acting on $\calR^4$, which may also give nonzero
contributions $\propto A_4$ in the \AdS5s5\ background.
There might also be terms that contribute to $A_4$
 that cannot be expressed in terms
of a local ten-dimensional action.

What about the possibility of terms which are non-perturbative in $\alpha'$,
which are  exponentially suppressed in
the Taylor expansion,
such as $e^{-{L^2/ \a}}$? For small $\a/ L^2$, we understand
terms of this type as coming from non-trivial saddle points of the world
sheet theory, namely world-sheet
instantons. But there are no topologically
non-trivial two-cycles for the world-sheet
to wrap in
$AdS_5\times S^5$ so such terms cannot appear in the
string genus expansion.\foot{ In other instances
of large $N$ gauge theories dual to closed strings,
world sheet instantons are present and play an important role \refs{\gopa}.}
One might argue that this does not rule out
$e^{-{L^2/\a}}$ contributions associated
with non-perturbative  $\tau$ dependence. But this also seems unlikely
since we understand such non-perturbative terms
as coming from D-instantons and again there is no obvious origin for
world-sheet instanton contributions in a D-instanton background.
We will therefore make the ansatz that
\taylexp\ is the complete expression for the coefficient of $A_4$
in the correlation function of four energy-momentum tensors,
at least for sufficently small $\a/L^2$.  In practice, we will only
need the
weaker assumption that this is so for the D-instanton contributions to
\taylexp\ (the terms with phases $e^{2\pi i K\tau_1}$, $|K|\ge 1$).
Our arguments will show that this assumption is, at
least, self consistent.

The $SL(2,Z)$ duality symmetry of the IIB theory is related via the AdS/CFT
correspondence to the Montonen--Olive duality of $\calN=4$ supersymmetric
Yang--Mills theory.  This requires that $F_l(\tau, {\bar \tau}, \eta)$
has specific  modular properties which means that it has the form,
\eqn\modprop{F_l(\tau, {\bar \tau}, \eta)= \t2^{1-{l\over 2}}
H_l(\tau, {\bar \tau}, \eta)}
where $H_l(\tau, {\bar \tau}, \eta)$ is modular invariant (a scalar
under $SL(2,Z)$). The explicit power
of $\tau_2^{1-{l\over 2}}$ arises  from the transformation in the effective action
from the string to the Einstein frame (where the metric is
$SL(2,Z)$ neutral)
due to the factors of the  metric appearing with the powers of $\alpha'/L^2$.
In terms of Yang-Mills variables, this is easy to see since
 the coefficient of $H_l$ in
\taylexp\
is the combination (from  \taylexp\ and \modprop)
\eqn\sltwoz{\left({\a\over L^2}
\right)^{l-2}\tau_2^{1-{l\over 2}}={1\over\l^{{l\over 2}-
1}}g_s^{{l\over 2}-1}= N^{-{l\over 2}+1},}
which is inert under $SL(2,Z)$.

Now consider the  non-perturbative part of the modular function
$H_l(\tau, {\bar \tau}, \eta)$
coming from  BPS charge-$K$ D-instantons.  This amounts to picking
out the  saddle point (when $g=\t2^{-1} <<1$)
with the exponential
$e^{2\pi (iK\tau_1 -|K|\tau_2)}$ dependence.\foot{In
the charge-$K$ sector with phase $e^{2\pi iK\tau_1}$
 there could be contributions from non-BPS configurations of
$K+K'$ instantons and $K'$ anti-instantons. These
would be suppressed by an additional factor of    $e^{-4\pi K' N/\l}$. Due to
their different $N$ dependence such terms would not enter our considerations,
even if they were present.}
We expect
this to take the generic form,
\eqn\hgen{H_l(\tau, {\bar \tau}, \eta)|_K =
d(K,l)e^{-2\pi(|K|\tau_2- iK\tau_1) }[h_0^{(l)}(\eta)+\tau_2^{-1}
h_1^{(l)}(\eta)+\tau_2^{-2}h_2^{(l)}(\eta)+ \ldots ],}
an expression which deserves further explanation.
 Firstly, the successive terms in
this series are spaced in integer powers of ${\tau_2}^{-1}$ since they
arise in string theory from world-sheet configurations
with increasing numbers of
boundaries with Dirichlet conditions and/or handles.
The functions $h_i^{(l)}(\eta)$ (which also depend on the instanton charge $K$),
that appear here,
are severely constrained by the
fact that $H_l$ is a modular function. Secondly, we have assumed that
there is no $\tau_2= g^{-1}$ dependence in the overall factor $d(K,l)$.
Any such  dependence would have to be a power that arises
 from the zero mode integrations around the D-instanton. This
power should not
depend on $l$. But we  know from the $l=1$ case  \fone\ that there is
no such overall factor.\foot{ Actually, the essence of our conclusion will not be
affected even if we had an overall factor of $g^{n_l}$, with $n_l$ taking
values over non-negative integers.}
This statement should also be a consequence of
supersymmetric cancellations of bosonic zero mode contributions with
fermionic ones.  Therefore, this might be special to a class of
 correlation functions such as the ones in the short multiplet we are
 concerned with.

Rewriting the amplitude in terms of
   $N$, $\l$ and $\theta$ and summing
the total contribution to the $K$ instanton background from all powers
in the derivative expansion (all $l$) gives (using \taylexp,\modprop\ and
\hgen)
\eqn\htot{\eqalign{\calH^{YM}|_K = &
 e^{-2\pi{|K|N\over \l} + iK\theta}\cr &
\sum_{l=1}d(K,l)N^{1-{l\over 2}}
\quad\left[h_0^{(l)}(\eta)+
\left({\l\over N}\right)
h_1^{(l)}(\eta)+\left({\l\over N}\right)^2h_2^{(l)}(\eta)+ \ldots
\right].
\cr}}
This can be reorganized into a 't Hooft expansion by
grouping  together the powers of $1/N$,  which gives the
perturbation expansion around the BPS $K$-instanton configuration,
\eqn\instexp{\calH^{YM}|_K =
 e^{-2\pi{|K|N\over \l}+iK\theta}\sum_{m=1}N^{1-{m\over 2}}f_m(\l,\eta).}
At first sight it  might seem surprising that the series of fluctuations
about an instanton should have a 't Hooft expansion in powers of
$N^{-1/2}$.  From the string viewpoint  the  spacing by half-integer
powers of $1/N$ arises very naturally
 from the presence of all integer powers of $\a$ in \taylexp.
Furthermore, this feature is confirmed directly in the Yang--Mills theory by
a saddle point analysis of the contribution of the exact zero-mode measure
(as for example in (5.7) of \refs{\doreythree}).  For general instanton number
$K>1$ the fluctuations around the saddle point are in  powers of $N^{-1/2}$.  The
series of
fractional powers of $N$ therefore arises from the $K$-instanton measure.
Only in the case  $K=1$ does the series consist of terms with integer
spaced powers of $1/N$ (starting with $N^{1/2}$).

The key point following from the structure of \htot\ is that the power of
$\l$ in the expansion is always bounded by that of $N$. In other words,
each function  $f_m(\l,\eta)$ in \instexp\  is
 a polynomial in $\l$,
\eqn\poly{f_m(\l,\eta)=\sum_{k=0}^{[{m-1\over 2}]}\l^{k}h^{(m-2k)}_{k}(\eta).}
For example,
\eqn\egf{f_1(\l,\eta)=h^{(1)}_{0}(\eta), \qquad
 f_2(\l,\eta)=  h^{(2)}_{0}(\eta), \qquad f_3(\l,\eta)= h^{(3)}_{0}(\eta)
+\l h^{(1)}_{1}(\eta), \qquad \dots.}
By our initial arguments, this is the complete form of the answer for
small $\a$ or equivalently large $\l$. Barring the (unlikely) possibility
of a phase transition as a function of $\l$, the knowledge that $f_m$ is
 a polynomial in positive powers of $\lambda$
 at large $\l$ allows it to be analytically continued to
weak 't Hooft coupling.\foot{Even
though the orginal expansion in $\a$ \taylexp\ and
$g$ \hgen\ may only be asymptotic, we can nevertheless trust
\poly\ to small $\l$. This is similar to the statement that though the full
perturbative expansion in gauge theories has zero radius of convergence,
the planar diagram expansion can be trusted in some finite radius \thf.}
But then the polynomial form of $f_m$
means that there are only  a finite number of terms in
the small-$\lambda$ perturbation expansion for each power of $N$.

Thus the very structure of the string expansion  implies a sequence of
non-trivial non-renormalisation theorems for 't Hooft perturbation theory
around an instanton background. In particular, the leading
large $N$ term comes from $m=1$ and that is just a constant as far as its
dependence on $\l$ is concerned. In other words, it receives only a
semi-classical contribution. This  `explains' the fact that the
semi-classical approximation to the $K$-instanton contribution to
$\calN=4$ $SU(N)$ Yang--Mills theory  at leading order in $N$
(the term of order $N^{1/2}$ which was evaluated
in  \refs{\doreythree})
agrees precisely with the expression predicted by
the AdS/CFT correspondence \refs{\bg}. Moreover, from \egf\ we see
that the next to leading term should behave as $N^{0}h_0^{(2)}(\eta)$ and
is also independent of $\lambda$ and therefore semi-classically exact.
Verifying this prediction would require knowledge of the $l=2$ term in
\taylexp\   that contributes
at order $(\a)^0$.

Other predictions can also be tested.  For instance,
the structure of $f_3$ in \egf\ requires that the next term  in
the $1/N$ expansion has only two terms in the  loop expansion.
Also, in the special case of instanton number $K=1$
the semi-classical contribution  was computed in
\doreytwo\ for all values of  $N$. In this case the expansion is
in integer powers of $1/N$, which immediately determines  that
\eqn\keqone{h_0^{(2k)}(\eta, K=1)=0, \qquad h_0^{(2k-1)}(\eta, K=1)=b_kg_1(\eta)}
where $b_k$ are the coefficients in the exact answer \refs{\doreytwo},
\eqn\expren{{\Gamma(N-{1\over 2})\over
\Gamma(N-1)} = N^{{1\over 2}} \sum_{l=1}^\infty b_k N^{-k+1} =
N^{{1\over 2}} \left(1 - {5\over 8}{1\over N} - {23 \over 128}
{1\over N^2} + \dots\right).}

\newsec{Comments and Conclusions}

We have seen by making rather minimal assumptions that
certain instanton contributions in the 't Hooft limit of
$\calN=4$ supersymmetric $SU(N)$ Yang--Mills theory receive only a finite
number of perturbative corrections at a given order in the $1/N$
expansion.  More precisely, at order $N^{1-{m\over 2}}$ there are
$[{m+1\over 2}]$ terms
in the power series  in the 't Hooft coupling, $\lambda$, starting with
$\lambda^0$.  It should be emphasised that, unlike with usual non-renormalisation
theorems, our statements only apply at each order in  $1/N$ in the 't Hooft limit
whereas for finite $N$
perturbative terms appear at all loops.  Although
our arguments do not make direct use of supersymmetry, this enters
indirectly since the  AdS/CFT correspondence does require  supersymmetry.
This is similar in spirit to the way in
which the mere existence of a Lorentz invariant eleven-dimensional
limit of  M-theory implies non-trivial facts about
D0-brane quantum mechanics, as in the
DLCQ description.

The assumptions we have made have the virtue that they
can be checked by direct evaluation of
perturbative contributions in the large-$N$ expansion of the gauge theory.
In this way one could investigate to what extent these results
apply to theories with
less supersymmetry.
It would, for instance, be interesting
to find out how much can be said about the conformal field
field theories described in \refs{\evashamit, \lnv}.
In such cases  the dual string theory has an $S^5/\Gamma$ sector
which could admit world sheet instantons whose absence was one of the
important ingredients in our argument.

\vskip0.3cm
{\it Acknowledgments}:  We would like to thank N. Berkovits, A. Lawrence,
J. Maldacena, S. Minwalla, H. Osborn,
A. Sen, A. Strominger and C. Vafa
for useful discussions. We would also like to acknowledge the stimulating
environment of the Cargese'99 workshop on Strings and M-theory as well as
the Strings '99 conference at Potsdam. R.G. would also like to thank
the Abdus Salam International Centre for Theoretical Physics
for it's hospitality during the
Extended  Workshop on String Theory and Dualities,
where part of this work was carried out.
The research of R.G. is supported by DOE grant DE-FG02-91 ER40654.

\listrefs

\end